\documentstyle[preprint,pre,aps,oldlfont]{revtex}
\begin{document}
\draft
%\preprint{KWNU-SP-94-4}
\title{ Extension of ``Renormalization of period doubling in symmetric
        four-dimensional volume-preserving maps''
}
\author{Sang-Yoon Kim \cite{byline}}
\address{
Department of Physics\\ Kangwon National University\\
Chunchon, Kangwon-Do 200-701, Korea
}
\maketitle
\begin{abstract}
We numerically reexamine the scaling behavior of period doublings
in four-dimensional volume-preserving maps in order to resolve a
discrepancy between numerical results on scaling of the
coupling parameter and the approximate renormalization results
reported by Mao and Greene [Phys. Rev. A {\bf 35}, 3911 (1987)].
In order to see the fine structure of period doublings, we extend the
simple one-term scaling law to a two-term scaling law. Thus we find a new
scaling factor associated with coupling and confirm
the approximate renormalization results.
\end{abstract}
%
%  END OF ABSTRACT
%

\narrowtext

Universal scaling behavior of period doubling has been found in
area-preserving maps
\cite{Benettin,Collet,Greene,Bountis,Helleman,Widom,MacKay}.
As a nonlinearity parameter is varied, an initially stable periodic orbit may
lose its stability and give rise to the birth of a stable period-doubled
orbit. An infinite sequence of such bifurcations accumulates at a finite
parameter value and exhibits a universal limiting behavior. However these
limiting scaling behaviors are different from those
for the one-dimensional dissipative case
\cite{Feigenbaum}.

An interesting question is whether the scaling results of
area-preserving maps carry over higher-dimensional volume-preserving maps.
Thus period doubling in four-dimensional (4D) volume-preserving
maps has been much studied in recent years
\cite{MacKay,Janssen2,Mao1,Mao2,Kim,Mao3}.
It has been found in Refs.~\cite{Mao2,Kim,Mao3} that the critical scaling
behaviors of period doublings for two symmetrically coupled area-preserving
maps are much richer than those for the uncoupled area-preserving case.
There exist an infinite number of critical points in the space of the
nonlinearity and coupling parameters. It has been numerically found in
\cite{Mao2,Kim} that the critical behaviors at those
critical points are characterized by two scaling factors, $\delta_1$ and
$\delta_2$. The value of $\delta_1$ associated with scaling of the
nonlinearity parameter is always the same as that of the
scaling factor $\delta$ $(=8.721\dots)$ for the area-preserving maps. However
the values of $\delta_2$ associated with scaling of the coupling parameter
vary depending on the type of bifurcation routes to the critical points.

The numerical results \cite{Mao2,Kim} agree
well with an approximate analytic renormalization results obtained by
Mao and Greene \cite{Mao3}, except for the zero-coupling case in which
the two area-preserving maps become uncoupled.
Using an approximate renormalization method including truncation, they
found three relevant eigenvalues, $\delta_1=8.9474$,
$\delta_2=-4.4510$ and $\delta_3=1.8762$ for the zero-coupling case
\cite{Mao4}. However  they believed that
the third one $\delta_3$ is an artifact of the truncation, because
only two relevant eigenvalues $\delta_1$ and $\delta_2$ could be
indentified with the scaling factors numerically found.

In this Brief Report we numerically study the critical behavior at the
zero-coupling point in two symmetrically coupled area-preserving maps
and resolve the discrepancy between the numerical
results on the scaling of the coupling parameter and the approximate
renormalization results for the zero-coupling
case. In order to see the fine structure of period doublings, we extend the
simple one-term scaling law to a two-term scaling law.
Thus we find a new
scaling factor $\delta_3=1.8505\dots$ associated with coupling, in
addition to the previously known coupling scaling factor $\delta_2
=-4.4038\dots$~.
The numerical values of $\delta_2$ and $\delta_3$ are close to the
renormalization results of the relevant coupling eigenvalues
$\delta_2$ and $\delta_3$.
Consequently the fixed map governing the critical behavior at the
zero-coupling point has two relevant coupling eigenvalues $\delta_2$ and
$\delta_3$ associated with coupling perturbations, unlike the cases of
other critical points.

Consider a 4D volume-preserving map $T$ consisting of two symmetrically
coupled area-preserving H${\acute {\rm e}}$non maps \cite{Mao2,Kim},
\begin{equation}
T:\left\{
       \begin{array}{l}
       x_1(t+1) = -y_1 (t) + f(x_1(t)) + g(x_1(t),x_2(t)), \\
       y_1(t+1) = x_1 (t), \\
       x_2(t+1) = -y_2 (t) + f(x_2(t)) + g(x_2(t),x_1(t)), \\
       y_2(t+1) = x_2 (t),
       \end{array}
   \right.
\label{eq:CM}
\end{equation}
where $t$ denotes a discrete time, $f$ is the nonlinear function of the
uncoupled H${\acute {\rm e}}$non's quadratic map \cite{Henon}, i.e.,
\begin{equation}
f(x) = 1- ax^2,
\label{eq:NF}
\end{equation}
and $g({x_1},{x_2})$ is a coupling function obeying a condition
\begin{equation}
g(x,x) = 0\;\;{\rm for\;\;any}\;\;x.
\label{eq:CC}
\end{equation}

The two-coupled map (\ref{eq:CM}) is called a symmetric map \cite{Mao2,Kim}
because it is invariant under an exchange of coordinates such that
$x_1 \leftrightarrow x_2$ and $y_1 \leftrightarrow y_2$.
The set of all points, which are invariant under the exchange of coordinates,
forms a symmetry plane on which $x_1=x_2$ and $y_1=y_2.$
An orbit is called an in-phase orbit if it lies on the symmetry plane,
i.e., it satisfies
\begin{equation}
x_1(t) = x_2(t) \equiv x(t), \;\;
y_1(t) = y_2(t) \equiv y(t)\;\; {\rm for \;\;all}\;t.
\end{equation}
Otherwise it is called an out-of-phase orbit.
Here we study only in-phase orbits. They can be easily found from the
uncoupled H${\acute {\rm e}}$non map because the coupling function $g$
satisfies the condition (\ref{eq:CC}).

Stability analysis of an in-phase orbit can be conveniently carried out
\cite{Mao2,Kim} in a set of new coordinates $(X_1,Y_1,X_2,Y_2)$ defined by
\begin{mathletters}
\begin{eqnarray}
X_1 &=& {(x_1+x_2) \over 2},\;\;\;  Y_1 = {(y_1+y_2) \over 2},\\
X_2 &=& {(x_1-x_2) \over 2},\;\;\;  Y_2 = {(y_1-y_2) \over 2}.
\label{eq:NC}
\end{eqnarray}
\end{mathletters}
Note that the in-phase orbit of the map (\ref{eq:CM}) becomes the orbit
of the new map (expressed in terms of new coordinates) with $X_2=Y_2=0$.
Moreover the new coordinates $X_1$ and $Y_1$ of the in-phase orbit also
satisfy the the uncoupled H${\acute {\rm e}}$non map.

Linearizing the new map at an in-phase orbit point, we obtain the
Jacobian
matrix $J$ which decomposes into two $2 \times 2$ matrices \cite{Mao2,Kim}:
\begin{equation}
{J} = \left ( \begin{array} {cc}
                   {J_1}& {\bf 0} \\
                   {\bf 0}& { J_2}
                   \end{array}
          \right ).
\label{eq:JM}
\end{equation}
Here $\bf 0$ is the $2 \times 2$ null matrix, and
\begin{eqnarray}
{J_1} &=&  \left ( \begin{array} {cc}
                   f'(X_1) & -1 \\
                   1 &    0
                     \end{array}
             \right ), \\
{J_2} &=&  \left ( \begin{array} {cc}
                   f'(X_1)-2 G(X_1) & -1 \\
                   1 &    0
                     \end{array}
             \right ),
\end{eqnarray}
where $f'(X) = {df \over dX}$ and $ \left. G(X) \equiv
{ {\partial g(X_1,X_2)} \over {\partial X_2}} \right|_{X_1=X_2=X}$.
Hereafter the function $G(X)$ will be called the ``reduced'' coupling
function of $g(X_1,X_2)$. Note also that the determinant of each $2 \times
2$ matrix $J_i$ $(i=1,2)$ is one, i.e., $Det({J_i})=1$.
Hence they are area-preserving maps.

Stability of an in-phase orbit with period $q$ is then determined
from the $q$-product $M_i$ of the $2 \times 2$ matrix
$ J_i$:
\begin{equation}
 {M_i} \equiv {\prod_{t=0}^{q-1}} {J_i}(X_1(t)), \;\;i=1,2.
\end{equation}
Since $Det(M_i)=1$, each matrix $M_i$ has a reciprocal
pair of eigenvalues, $\lambda_i$ and $\lambda_i^{-1}$.
Associate with a pair of eigenvalues $(\lambda_i,\lambda_i^{-1})$
a stability index \cite{Howard},
\begin{equation}
\rho_i = \lambda_i + \lambda_i^{-1},\;\;i=1,2,
\end{equation}
which is just the trace of $M_i$, i.e., $\rho_i=Tr(M_i)$.
Since $M_i$ is a real matrix, $\rho_i$ is always real.
Note that the first stability index $\rho_1$ is just that for the case of
the uncoupled H${\acute {\rm e}}$non map and hence coupling affects only
the second stability index $\rho_2$.

An in-phase orbit is stable only when the moduli of its stability indices
are less than or equal to two, i.e., $|\rho_i| \leq 2$ for $i=1$ and $2$.
A period-doubling (tangent) bifurcation occurs when each
stability index $\rho_i$ decreases (increases) through $-2$ (2).
Hence the stable region of the in-phase orbit in the parameter plane is
bounded by four bifurcation lines associated with tangent and
period-doubling bifurcations (i.e., those curves determined
by the equations $\rho_i = \pm 2$ for $i=0,1$).
When the stability index $\rho_1$ decreases
through $-2$, the in-phase orbit loses its stability via in-phase
period-doubling bifurcation and gives rise to the birth of the period-doubled
in-phase orbit. Here we are interested in scaling behaviors of such in-phase
period-doubling bifurcations.

As an example we consider a linearly-coupled case in which the coupling
function is
\begin{equation}
g(x_1,x_2)= {c \over 2} (x_2 - x_1).
\label{eq:CF}
\end{equation}
Here $c$ is a coupling parameter.
As previously observed in Refs. \cite{Mao2,Kim}, each ``mother'' stability
region bifurcates into two ``daughter''stability regions successively in the
parameter plane. Thus the stable regions of in-phase orbits of period
$2^n$ $(n=0,1,2,\cdots)$ form a ``bifurcation'' tree in the parameter plane
\cite{Rem}.

An infinite sequence of connected stablity branches (with increasing period)
in the bifurcation tree
is called a bifurcation ``route'' \cite{Mao2,Kim}.
Each bifurcation route can be represented by its address, which is an
infinite sequence of two symbols (e.g.,  $L$ and $R$).
A ``self-similar'' bifurcation ``path'' in a bifurcation route is formed
by following a sequence of parameters $(a_n,c_n)$, at which the in-phase
orbit of level $n$ (period $2^n$) has some given stability indices
$(\rho_1, \rho_2)$ (e.g., $\rho_1=-2$ and $\rho_2=2$) \cite{Mao2,Kim}.
All bifurcation paths within a bifurcation route  converge to an
accumulation point $(a^*,c^*)$, where the value of $a^*$ is always the
same as that of the accumulation point for the area-preserving case
(i.e., $a^*=4.136\,166\,803\,904 \dots$), but
the value of $c^*$ varies depending on the bifurcation routes.
Thus each bifurcation route ends at a
critical point $(a^*, c^*)$ in the parameter plane.

It has been numerically found that scaling behaviors near a critical point
are characterized by two scaling factors, $\delta_1$ and $\delta_2$
\cite{Mao2,Kim}. The value of $\delta_1$ associated with scaling of the
nonlinearity parameter is always the same as that of the scaling factor
$\delta$ $(=8.721\dots)$ for the area-preserving case. However the values
of $\delta_2$ associated with scaling of the coupling parameter vary
depending on the type of bifurcation routes. These
numerical results agree well with analytic renormalization results
\cite{Mao3}, except for the case of one specific
bifurcation route, called the $E$ route. The address of the $E$ route is
$[(L,R,)^\infty]$ $(\equiv [L,R,L,R,\dots])$ and it ends at the
zero-coupling critical point $(a^*,0)$.

Using an approximate renormalization method including truncation, Mao and
Greene \cite{Mao3} obtained three relevant eigenvalues, $\delta_1 = 8.9474$,
$\delta_2 = -4.4510$, and $\delta_3 = 1.8762$ for the zero-coupling case;
hereafter the two eigenvalues $\delta_2$ and $\delta_3$ associated with
coupling will be called the coupling eigenvalues (CE's).
The two eigenvalues $\delta_1$ and $\delta_2$ are close to the numerical
results of the nonlinearity-parameter scaling factor $\delta_1(=8.721\dots)$
and the coupling-parameter scaling factor $\delta_2(=-4.403\dots)$ for the
$E$ route.
However they believed that the second relevant CE $\delta_3$ is an
artifact of the truncation, because it could not be identified with
anything obtained by a direct numerical method.

In order to resolve the discrepancy between the numerical results and
the renormalization results for the zero-coupling case, we numerically
reexamine the scaling behavior associated with coupling.
Extending the simple one-term scaling law to a two-term scaling law, we
find a new scaling factor
$\delta_3 = 1.8505\dots$ associated with coupling in addition to the
previously found coupling scaling factor $\delta_2=-4.4038\dots$, as will
be seen below. The values of these two coupling scaling factors are close
to the renormalization results of the relevant CE's $\delta_2$ and
$\delta_3$.

We follow the in-phase orbits of period $2^n$ up to level $n=14$
in the $E$ route and
obtain a self-similar sequence of parameters $(a_n,c_n)$, at which
the pair of stability indices, $(\rho_{0,n},\rho_{1,n})$, of the orbit
of level $n$ is $(-2,2)$.
The scalar sequences $\{ a_n \}$ and $\{ c_n \}$ converge  geometrically
to their limit values, $a^*$ and $0$, respectively.
In order to see their convergence, define
$\displaystyle{ \delta_n \equiv {\Delta a_{n+1} \over \Delta a_n}}$ and
$\displaystyle{ \mu_n \equiv {\Delta c_{n+1} \over \Delta c_n}},$
where $\Delta a_n = a_n - a_{n-1}$ and
$\Delta c_n = c_n - c_{n-1}$.
Then they converge to their limit values $\delta$ and $\mu$ as $n \rightarrow
\infty$, respectively. Hence the two sequences $\{ \Delta a_n \}$ and
$\{ \Delta c_n \}$ obey one-term scaling laws asymptotically:
\begin{equation}
\Delta a_n = C^{(a)} \delta^{-n},\;\;\;
\Delta c_n = C^{(c)} \mu^{-n}\;\;\;{\rm for\;large\;}n,
\label{eq:OTSL}
\end{equation}
where  $C^{(a)}$ and $C^{(c)}$ are some constants, $\delta=8.721 \cdots$,
and $\mu=-4.403 \cdots$. The values of $\delta$ and $\mu$ are close to the
renormalization results of the first and second relevant eigenvalues
$\delta_1$ and $\delta_2$, respectively.

In order to take into account the effect of the second relevant CE
$\delta_3$ on the scaling of the sequence $\{ \Delta c_n \}$,
we extend the simple one-term scaling law (\ref{eq:OTSL}) to a two-term
scaling law:
\begin{equation}
\Delta c_n = C_1 \mu_{1}^{-n} + C_2 \mu_{2}^{-n} \;\;\;{\rm for\;large\;}n,
\label{eq:TTSL1}
\end{equation}
where $| \mu_1 | > | \mu_2 |$.
This is a kind of multiple scaling law \cite{MR}.
Eq.~(\ref{eq:TTSL1}) gives
\begin{equation}
\Delta c_n = t_1 \Delta c_{n+1} -t_2 \Delta c_{n+2},
\label{eq:RE}
\end{equation}
where $t_1 = \mu_1 + \mu_2$ and $t_2 = \mu_1 \mu_2$.
Then $\mu_1$ and $\mu_2$ are solutions of the following quadratic equation,
\begin{equation}
\mu^2 - t_1 \mu + t_2 =0.
\label{eq:QE}
\end{equation}
To evaluate $\mu_1$ and $\mu_2$, we first obtain $t_1$ and $t_2$ from
$\Delta c_n$'s using Eq.~(\ref{eq:RE}):
\begin{mathletters}
\label{eq:T1T2}
\begin{eqnarray}
t_1 &=& { {\Delta c_n \Delta c_{n+1} - \Delta c_{n-1} \Delta c_{n+2}}
\over {\Delta c_{n+1}^2 - \Delta c_n \Delta c_{n+2}} }, \\
t_2 &=& { {\Delta c_n^2 - \Delta c_{n+1} \Delta c_{n-1}}
\over {\Delta c_{n+1}^2 - \Delta c_n \Delta c_{n+2}} }.
\end{eqnarray}
\end{mathletters}
Note that Eqs.~(\ref{eq:TTSL1})-(\ref{eq:T1T2}) hold only for large $n$.
In fact the values of $t_i$'s and $\mu_i$'s $(i=1,2)$ depend on the
level $n$. Therefore we explicitly denote $t_i$'s and $\mu_i$'s by
$t_{i,n}$'s and $\mu_{i,n}$'s, respectively. Then each of them converges
to a constant as $n \rightarrow \infty$:
\begin{equation}
\lim_{n \rightarrow \infty} t_{i,n} = t_i, \;\;\;
\lim_{n \rightarrow \infty} \mu_{i,n} = \mu_i,\;\;i=1,2.
\end{equation}

Three sequences $\{ \mu_{1,n} \}$, $\{ \mu_{2,n} \}$, and
$\displaystyle{ \{ {\mu_{1,n}^2 / \mu_{2,n}} \} }$ are shown in
Table \ref{table1}.
The second column shows rapid convergence of $\mu_{1,n}$ to its limit
values $\mu_1$ $(=-4.403\,897\,805)$, which is close to
the renormalization result of the first relevant CE (i.e.,
$\delta_2$ $=-4.4510$).
{}From the third and fourth columns, we also find that the second scaling
factor $\mu_2$ is given by a product of two relevant CE's $\delta_2$ and
$\delta_3$,
\begin{equation}
\mu_2 = {\delta_2^2 \over \delta_3},
\end{equation}
where $\delta_2=\mu_1$ and $\delta_3=1.850\,65$~.
It has been known that every scaling factor in the multiple-scaling
expansion of a parameter is expressed by a product of the eigenvalues of a
linearized renormalization operator \cite{MR}.
Note that the value of $\delta_3$ is close to the renormalization result
of the second relevant CE (i.e., $\delta_3=1.8762$).

We now study the coupling effect on the second stability index $\rho_{2,n}$
of the in-phase orbit of period $2^n$ near the
zero-coupling critical point $(a^* , 0)$.
Figure \ref{figure1} shows three plots of $\rho _{2,n} (a^*,c)$ versus
$c$ for $n=4,5,$ and $6$. For $c=0$,
$\rho_{2,n}$ converges to a constant $\rho_2^*$ $(=-2.543\,510\,20\dots)$,
called the critical stability index \cite{Kim}, as $n \rightarrow \infty$.
However, when $c$ is non-zero $\rho_{2,n}$ diverges as $n \rightarrow
\infty$, i.e., its slope $S_n$
$\displaystyle{ \left.
(\equiv  {{\partial \rho_{2,n}} \over {\partial c}} \right|_{(a^* ,0)})
}$
at the zero-coupling critical point
diverges as $n \rightarrow \infty$.

The sequence $\{ S_n \}$ obeys a two-term scaling law,
\begin{equation}
S_n = D_1 \nu_1 ^n + D_2 \nu_2 ^n\;\;\;{\rm for\;large\;}n,
\label{eq:TTSL2}
\end{equation}
where $|\nu_1| > |\nu_2|$.
This equation gives
\begin{equation}
S_{n+2} = r_1 S_{n+1} - r_2 S_{n},
\end{equation}
where $r_1 = \nu_1 + \nu_2$ and $r_2 = \nu_1 \nu_2$.
As in the scaling for the coupling parameter, we first obtain
$r_1$ and $r_2$ of level $n$ from $S_n$'s:
\begin{equation}
r_{1,n} = {  {S_{n+1} S_{n} - S_{n+2} S_{n-1}}
\over {S_{n}^2 - S_{n+1} S_{n-1}} },\;\;\;
r_{2,n} = { { S_{n+1}^2 - S_{n} S_{n+2}}
\over {S_{n}^2 - S_{n+1} S_{n-1}} }.
\end{equation}
Then the scaling factors $\nu_{1,n}$ and $\nu_{2,n}$ of level $n$ are given
by the roots of the quadratic equation, $\nu_n^2 - {r_{1,n}} {\nu_n} +
{r_{2,n}} =0$. They are listed in Table \ref{table2} and converge to
constants $\nu_1$ $(=-4.403\,897\,805\,09)$ and $\nu_2$ $(=1.850\,535)$
as $n \rightarrow \infty$, whose accuracies are higher than those of the
coupling-parameter scaling factors.
Note that the values of $\nu_1$ and $\nu_2$ are
also close to the renormalization results of the two relevant CE's $\delta_2$
and $\delta_3$.

We have also studied several other coupling cases with the coupling function,
$g(x_1 , x_2) = {\displaystyle {c \over 2}} (x_2^n - x_1^n) $
($n$ is a positive integer).
In all cases studied $(n=2,3,4,5)$, the scaling factors of both the coupling
parameter $c$ and the slope of the second stability index $\rho_2$ are found
to be the same as those for the above linearly-coupled case $(n=1)$ within
numerical accuracy. Hence universality also seems to be well obeyed.

\acknowledgments

This work was supported by the Basic Science Research Institute Program,
Ministry of Education, Korea, 1994, Project No. BSRI-94-2401.

%
% END OF REFERENCES
%

\begin{table}
\caption{ Scaling factors $\mu_{1,n}$ and $\mu_{2,n}$ in the two-term
          scaling for the coupling parameter are shown in the second
          and third columns, respectively. A product of them,
          $\displaystyle{ {\mu_{1,n}^2 \over \mu_{2,n}} }$, is shown in the
          fourth column.}
\begin{tabular}{cccc}
$n$ & $\mu_{1,n}$ & $\mu_{2,n}$ &
$\displaystyle{ {\mu_{1,n}^2 \over \mu_{2,n}} }$ \\
\tableline
5 & -4.403\,908\,128 & 10.437\,4 & 1.858\,17 \\
6 & -4.403\,899\,694 & 10.465\,9 & 1.853\,09 \\
7 & -4.403\,898\,736 & 10.458\,2 & 1.854\,46 \\
8 & -4.403\,897\,867 & 10.474\,8 & 1.851\,52 \\
9 & -4.403\,897\,847 & 10.473\,9 & 1.851\,68 \\
10& -4.403\,897\,806 & 10.478\,4 & 1.850\,89 \\
11& -4.403\,897\,807 & 10.478\,6 & 1.850\,85 \\
12& -4.403\,897\,805 & 10.479\,7 & 1.850\,65 \\
\end{tabular}
\label{table1}
\end{table}

\begin{table}
\caption{ Scaling factors $\nu_{1,n}$ and $\nu_{2,n}$ in the two-term
          scaling for the slope of the second stability index are shown.}
\begin{tabular}{ccc}
$n$ & $\nu_{1,n}$ & $\nu_{2,n}$ \\
\tableline
5 & -4.403\,898\,453\,59 &  1.851\,433\,5 \\
6 & -4.403\,897\,730\,29 &  1.850\,782\,6 \\
7 & -4.403\,897\,813\,85 &  1.850\,603\,6 \\
8 & -4.403\,897\,804\,07 &  1.850\,553\,8 \\
9 & -4.403\,897\,805\,21 &  1.850\,540\,0 \\
10& -4.403\,897\,805\,07 &  1.850\,536\,1 \\
11& -4.403\,897\,805\,09 &  1.850\,535\,0 \\
12& -4.403\,897\,805\,09 &  1.850\,534\,9 \\
\end{tabular}
\label{table2}
\end{table}
%
%  End of Tables
%
\begin{figure}
\caption{Plots of the second stability index $\rho_{2,n} (a^*,c)$ versus $c$
        for $n=4,5,6$. ~~~~~~~~~~~~~~~~~~~~~~~~~~~}
\label{figure1}
\end{figure}
%
%  End of Figures Captions
%


\begin{references}
\bibitem[*]{byline} Electronic Address: sykim@cc.kangwon.ac.kr (Internet)
\bibitem{Benettin} G.\ Benettin, C.\ Cercignani, L.\ Galgani, and
 A.\ Giorgilli, Lett.\ Nuovo Cimento {28}, 1 (1980); {\bf 29}, 163 (1980).
\bibitem{Collet} P.\ Collet, J.\ P.\ Eckmann, and H.\ Koch, Physica D
{\bf 3}, 457 (1981).
\bibitem{Greene} J.\ M.\ Greene, R.\ S.\ MacKay, F.\ Vivaldi, and M.\
J.\ Feigenbaum, Physica D {\bf 3}, 468 (1981).
\bibitem{Bountis} T.\ C.\ Bountis, Physica D {\bf 3}, 577 (1981).
\bibitem{Helleman} R.\ H.\ G.\ Helleman, in {\it Long-Time Prediction in
 Dynamics}, edited by W.\ Horton, L. Reichl, and V. Szebehely (Wiley, New
 York, 1982), pp.\ 95-126.
\bibitem{Widom} M.\ Widom and L.\ P.\ Kadanoff, Physica D {\bf 5}, 287 (1982).
\bibitem{MacKay} R.\ S.\ MacKay, Ph.D. Thesis, Princeton University, 1982.
\bibitem{Feigenbaum} M.\ J.\ Feigenbaum, J.\ Stat.\ Phys.\ {\bf 19},25 (1978);
{\bf 21}, 669 (1979).
\bibitem{Janssen2} T.\ Janssen and J.\ A.\ Tjon, J.\ Phys.\ A {\bf 16},
 673 (1983); {\bf 16}, 697 (1983).
\bibitem{Mao1} J.-m.\ Mao, I.\ Satija, and B.\ Hu, Phys.\ Rev.\ A {\bf 32},
 1927 (1985); {\bf 34}, 4325 (1986).
\bibitem{Mao2} J.-m. Mao and R.\ H.\ G.\ Helleman, Phys.\ Rev.\ A {\bf 35},
 1847 (1987).
\bibitem{Kim} S.-Y.\ Kim and B.\ Hu, Phys.\ Rev.\ A {\bf 41}, 5431 (1990).
\bibitem{Mao3} J.-m. Mao and J.\ M.\ Greene, Phys.\ Rev.\ A {\bf 35}, 3911
  (1987)
\bibitem{Mao4} See the table I in Ref.~\cite{Mao3}. The $\delta_3$ in the
text corresponds to $\delta_2'$ in the table.
\bibitem{Henon} M.\ H${\acute {\rm e}}$non, Quart.\ Appl.\ Math.\ {\bf 27},
 291 (1969).
\bibitem{Howard} J.\ E.\ Howard and R.\ S.\ MacKay, J.\ Math.\ Phys.\
 {\bf 28}, 1036 (1987).
\bibitem{Rem} A ``bifurcation'' tree in the parameter plane
               is shown in the figure 1 in each of Refs.~\cite{Mao2} and
               \cite{Kim}.
\bibitem{MR} J.-m Mao and B. Hu, J. Stat. Phys. {\bf 46}, 111 (1987);
 Int. J. Mod. Phys. B {\bf 2}, 65 (1988); C. Reick, Phys. Rev. A {\bf 45},
 777 (1992).
\end{references}
\end{document}